\documentclass[twocolumn,showpacs,preprintnumbers,amsmath,amssymb,floatfix]{revtex4}

\usepackage{graphicx}
\usepackage{dcolumn}
\usepackage{bm}

\begin{document}
\preprint{PNPI/08.2006}

\title {Currents of non-uniformities in solar atmosphere}
\author{ S.~I.~Stepanov }
\altaffiliation{Petersburg Nuclear Physics Institute of the Russian Academy of Sciences, St.-Petersburg,  Russia}
\email{stepanovsmail@mail.ru}
\date{\today}

\begin{abstract}
Non-uniformities of plasma and magnetic field are known to cause
electric currents in plasma. Electron density gradient causes
diffusion current, electron temperature gradient --- thermocurrent, gradient
of magnetic field module --- gradient current, curvature of magnetic field
lines --- centrifugal current. Being independent of electric field, the
currents of non-uniformities may act as extraneous to cause charge
separation and electric field in plasma. In cosmos,
the currents of non-uniformities were observed; in
particular, gradient and centrifugal currents --- in
magnetosphere, diffusion one --- in a comet coma and in artificial
plasma cloud.
On present work, the gradient current was investigated more fully than
earlier. Two unknown components, parallel and perpendicular to magnetic field
were found. The equation for gradient current density was obtained.
We compared the theoretical densities of currents of non-uniformities
(with usage of electron pressure and corresponding gradients) with
measured current densities (calculated as rotor of magnetic
field) for sun photosphere.
It follows from the comparison that the
currents of non-uniformities play important, may be main, role in
measured local current in photosphere. It is necessary to consider
in electromagnetic models.
\end{abstract}
\maketitle
\subsection*{Introduction}

It is commonly believed that magnetic field of Sun is generated by
convective motions of plasma inside the star. The field, together with
the electric currents being the source of the field, rises to upper
layers of Sun, its atmosphere. There, the field gives birth the
observed phenomena of solar activity such as spots, active regions,
flares, and mass ejections. The region of generation and the region of observation very
differ on the plasma parameters, in particular on plasma density.
This leads to a difference in the values of electron magnetization,
$\beta=\omega / \nu $.
(Here $\omega$ is the gyromagnetic frequency, $\nu$  is the frequency of
isotropization collisions.) In convective zone of the Sun (region of generation)
$\beta \approx 0$, at the photosphere level $\beta \approx 1$, at chromosphere $\beta$
amounts to hundreds,
and it is still larger in corona. Because of very different magnetization
one should apply different approaches for modeling electrical processes in these
media. Inside the Sun the conductivity may be regarded as scalar;
the conductivity current be collinear to electric field. This is namely
the basis of standard magmetohydrodynamics (MHD) models. In the star atmosphere the plasma is
anisotropic. In this case, the conductivity current has tree
components, --- parallel (magnetic field-aligned) current, Hall current, and
perpendicular (Pederson) current. It is obvious that the standard MHD model
being applied to the anisotropic plasma of star atmosphere cannot bring a true result.
To take into account the anisotropy, the simplest approach now is used:
the current is assumed to be parallel
to the magnetic field (free-force approach).

Plasma of solar atmosphere is very non-uniform medium; this appears in
gradients of density, temperature and non-uniform magnetic fields.
In such the plasma, currents of non-uniformities --- diffusion, thermocurrent, and gradient current
--- may play an essential role.
At present we may see the tendency to take partly
into account these currents.
Electromagnetic plasma models have been published with having applied
Hall current~\cite{PR_61:N836,PRL_93:N1062}, diffusion current~\cite{PhScr_57:N892},
thermocurrent~\cite{PRL_N:726}, and with applying a number of
currents~\cite{ApJ_1034:509,SP_N:1084}.

Distribution of magnetic fields and currents in solar atmosphere
is a hot point of solar physics.
The current density in photosphere may be deduced by means of magnetic
field rotor; or the presence of the currents may be found on deviation
of the field from so-called potential field. Structure of magnetic field
in solar atmosphere may be revealed on soft X-rays and H$\alpha$ images,
and on extrapolation of the field being measured on photosphere level.

It is known that vertical magnetic field-aligned currents with the value
up to $10^{12}$~A flow upward and downward in a sunspot~\cite{Severny1965}.
It turned out that these currents, being close together, are an universal
property of the spot, and moreover they cover a large share of an active
region~\cite{SP_N:900Abram,AA_N:1057,SP_N:1081,SP_N:1069,AA386:646,SP174:175,AA392:1119}.
A little is known on nature of the currents. In electromagnetic models
they are simply postulated as initial conditions~\cite{AA_337,AA_351:Titov}.

It was found that
twisted magnetic flux tubes represent typical phenomenon. Another
observation relatively the tubes --- they expand slightly with increasing
the height in atmosphere~\cite{AA_N:1057,SP_N:1081,SP_N:1069,AA_336:359,ApJ_557:880}.
It follows from Maxwell equations that both the phenomena arise
because of electric currents being present in the tube. The twist
arises due to longitudinal currents, the absence of tube expanding
appears due to currents embracing the tube, --- this is akin to
a solenoid current encircling the magnetic field of constant diameter.

It is difficult to study horizontal currents by a direct method,
by means of calculating the magnetic field rotor. Sparse papers
show that these currents take place in active regions~\cite{SP_N1071}.
Accurate analysis of magnetic field strength at some heights over photosphere
was fulfilled in~\cite{SP_N1077}. The field in temperature minimum zone was found
being noticeably larger than in middle photosphere. This points out at existence of
horizontal currents in there.

Studying the currents in active regions is of particular interest in the
context of flare activity. There are theoretical and experimental evidences
that solar flares are energized by the magnetic field in active
region. At one of first theory~\cite{Jacobsen,SP_1:Alfven}, a loop of
electric current plays main role in a flare. A part of the loop is situated
beneath photosphere, and another part --- in solar atmosphere. The loop carries
the current up to $10^{12}$~A. The flare begins when suddenly disruption of the
current happens in atmospheric part of the loop. In the place of disruption
a large electrical voltage, say $10^{10}$~V, appears. On theory, the place of energy
release has the dissipation factor ${\bf j} \cdot {\bf E}$ (a scalar product) being
positive. It is important that the place of energy release has very smaller volume
than the place where the magnetic energy was stored.

In general, such the scenario proves be true in later observations.
The sites of flares are linked with the places where the currents are
concentrated at photosphere level, in other words, --- with the places of
non-potential magnetic field~\cite{SP_N1077,SP_N1065,SP_N1080}. Some studies
show that flare is coupled with two magnetic field loops~\cite{AA_N:1057,SP_N1066,AA_344:981}
or with several loops from which only two play main role~\cite{SP_N:1069,AA_336:359}.
Both loops carry electric currents, in particular parallel currents. One, a
high-lying loop, exists long before the flare. Certainly this loop accumulates
magnetic energy. The onset of the flare correlates with the emergence of a smaller
low-lying loop. There are observations that the flare begins in the place where
the loops interact --- X-ray source may be seen here. Then the flare process spreads
along the loops attaining their footpoints. The flare energy is about
5--10\% of stored magnetic energy; the high-lying loop is seen to survive
after the flare. Maximal magnetic field strength was noticed in peak of
the flare~\cite{SP_N:1081,SP_N1077}.
The magnetic fields and electric currents often appear to simplify their structure
after the flare~\cite{AA_334:L57}.

Magnetic fields are everywhere observed in galaxies. In spirals,
orientation of the field are often (not always) parallel to arms: the field is of the values
3--10~$\mu G$~\cite{AA_348:405}. As mentioned for a long time, magnetization is
very high for all kinds of interstellar medium. Electron magnetization in cold
interstellar gas, zone H1, ($T = 30-70^{\circ}K$, hydrogen concentration is of 20--40~$cm^{-3}$~)
amounts to $10^7$, electron magnetization in coronal ionized gas is of
$10^{11}$~\cite{Parker:Book}.
Magnetic field with the strength of 22~$m G$ was found in the corona of the circumstellar
disk of young stellar object at the distance 40 a.u. from the object~\cite{AA_344:923}.
Even in that, most dense medium, magnetization is not smaller than 200.

In contrast to stars, magnetic field in intragalactic medium is observed in
the same place where being generated. Because the magnetization is very high there,
the electromagnetic theory should be from the very outset formulated on base of
anisotropic conductivity. This relates fully to the magnetic field in clusters
of galaxies.

In Earth magnetosphere, at the latitude of some ten thousand kilometers, a
ring current flows; it was directly detected in flyby of rocket~\cite{DAN_N464}.
Magnetic storm arises when a cloud of coronal plasma enters inside Earth's
magnetic field and this causes increase in this magnetosphere current. It is
well known that the magnetosphere current represents, on its nature, a sum of
gradient and centrifugal currents. Notice, the magnetization in the
magnetosphere is very high. The same conditions --- the gradient of magnetic
field module, the curvature of magnetic lines, and high magnetization take place
in stellar atmosphere. Hence, one should expect that the gradient and centrifugal
currents are present there; and the task arises --- how much large are the currents
and what is their role in electromagnetic phenomena?

\subsection*{Plasma currents arising from non-uniformities}

A current of charged particles arises due to fields, non-uniformity, and anisotropy
of medium~\cite{UFN_N220}. Let us list the currents what were detected in cosmic plasmas:

1. Conductivity current. It is caused by electric field.

2. Diffusion current. It is due to gradient of charged particle density.

3. Thermocurrent. This current is also named as "thermodiffusion" current,
"thermoelectric effect" and "Nernst effect"~\cite{PRL_N:726}. This current is
due to gradient of charged particle temperature provided that collision frequency
has some dependence of velocity of charged particles~\cite{N971:Tzen}. In the absence
of magnetic field the nature of diffusion and thermocurrents are quite similar to
diffusion and thermodiffusion in gas. In presence of magnetic field both the
currents (more precisely their Hall components) are often called as "diamagnetic effect" because
this effect leads to weakening of magnetic field in spatially restricted plasma.
In cosmic conditions diamagnetic effect was confidently observed in comet
coma~\cite{AA_N1052} and in the plasma cloud made by explosion of metal in
ionosphere~\cite{GRL_N786}.

4. Gradient current. It arises when the module of magnetic field changes in space.

5. Centrifugal current. It arises in magnetic field with curved
lines.

These currents arise due to non-uniformities of potential, plasma,
and magnetic field. We will name these currents as the currents of
non-uniformities.

In some cases, another currents may be essential, for example, the current
of entrainment of electrons by photons and currents caused by waves.

A current in plasma is calculated in different ways depending of the parameter
$\lambda /l$, $\lambda $  is the free path (or giroradius) of the particle,
$l$  is the typical plasma dimension. For the case of finite $\lambda / l$,
each the current must be calculated for given plasma geometry, magnetic field,
etc. For tokomaks, the current dependent of plasma and temperature gradients,
--- named "bootstrap current", has been calculated~\cite{PRL_93:N1047}. In cosmic plasma
the relation  $\lambda /l \approx 0$ takes place as a rule.
In this case the currents are roughly defined by local plasma conditions and universal
equations may be obtained for density of different  currents.
But in some cases the currents are non-local:
the current associated with particles accelerated in flares, for example.

The total current is the vector
sum of currents of the different kinds, each for electrons and ions:
\begin{eqnarray}
\lefteqn{{\bf j}={\bf j}_{E}+{\bf j}_{D}+{\bf j}_{T}+{\bf j}_{\nabla
B}+{\bf j}_{R}+} \nonumber\\
&& + \, \mbox{similar currents for ions} \, + \, \mbox{another currents}.
\end{eqnarray}
Here ${\bf j}_{E}$, ${\bf j}_{D}$, ${\bf j}_{T}$, ${\bf j}_{\nabla B}$, and ${\bf j}_{R}$
designates the density of conductivity,
diffusion, thermocurrent, gradient, and centrifugal current, accordingly.
It is natural, that the density of the total current
simultaneously satisfies to Maxwell equation thereby making it possible
measuring the current through magnetic field rotor.

From the available literature it appears that equations  for
density of the conductivity, diffusion, and thermocurrents
are known for a wide range of plasma parameters~\cite{N971:Tzen,Raizer}.
At the presence of a magnetic field each the current has three
components. For example, components of the conductivity current are
directed on the mutually perpendicular vectors ${\bf E_{\parallel}}$,
${\bf E_{\perp} }$, and ${\bf E} \! \times \! {\bf B }$ --- parallel, perpendicular,
and Hall component, respectively.
The components of diffusion and thermocurrent have same names.

The gradient of magnetic field module, $\nabla \! B $ , is an axial vector.
Generally, it can be oriented with any way concerning the
magnetic field in the point. This allows decomposing the
gradient into two components $\nabla_{\parallel} B$ and $\nabla_{\perp} B$,
being parallel and perpendicular to the magnetic field, Fig.~\ref{FigCoord}(a).
The mutually perpendicular vectors
$\nabla_{\parallel} B $, $\nabla_{\perp} B$,
and $ {\bf B }\times \nabla \! B$  define the coordinate axes Z, X,
and Y, respectively. When speaking about the gradient current,
one keeps in mind its component, which being directed at the vector
$ {\bf B } \times \nabla \! B$,
see Ref.~\cite{Alfven:book}, for example. The formula for this "Hall"
component is deduced and applied for a case of infinite
magnetization  --- collisionless plasma. The origin of this
component is presented on Fig.~\ref{FigHall}(b). The Hall gradient current is
well known on observations of electric phenomena in plasmas.
As mention above, this current creates, together with the
centrifugal current, the ring current in magnetosphere.
It is likely assume, that there are not one, but
three components of the gradient current.
Two unknown components will be considered as directed on the vectors
$\nabla_{\parallel} B$ and $\nabla_{\perp} B$
and be named as parallel and perpendicular components, respectively.

\subsection*{Parallel component of gradient current}

Let us show that the parallel component of the gradient current does exist,
and estimate its value. In the beginning
of the consideration we will follow to historically developed scheme,
for example, see~\cite{Alfven:book}.

In magnetic field a charged particle gyrates.
This movement may be characterized by a magnetic moment
\begin{equation}
\mu = m V^{2}_{\perp} / 2  B = \varepsilon_{\perp}/ B.
\end{equation}
Here $V_{\perp}$  is the perpendicular (to the field) component of velocity,
$\varepsilon_{\perp}$  is the perpendicular kinetic energy, $m$ is the mass
of the particle.

For researching the parallel component it is convenient to take
the field in which the gradient is parallel to
the field itself, Fig.~\ref{FigCoord}(b) and Fig.~\ref{FigParal}. As known, in this field
the particle undergoes to the mean Lorentz force
\begin{equation}
\langle f \rangle = - \mu \nabla \! B.
\end{equation}
Irrespective of the sign of the charge the force is directed to weaker field,
see Fig.~\ref{FigParal}.

As the particle gyrates, the parallel component of velocity increases,
and the perpendicular one decreases. Total velocity of the particle
remains, because kinetic energy of the particle does not change
in interaction with a magnetic field. Thus there is ejection of the charged
particles and thereby all the plasma into the weaker field area.
The ejection of plasma is essential
for very and fully ionized plasma, --- this
is chromosphere and corona as applied to the Sun.
The magnetic field with $ \nabla \! B \parallel \! {\bf B}$
is used in the devices for magnetic confinement of plasma ---
magnetic bottles. Here we complete citing the historical consideration.

Let's continue the consideration to show that the longitudinal
Lorentz force causes an electric current. We will consider weakly
ionized plasma, i.e., in which the electron-neutral and ion-neutral
collisions prevail over the collisions between the charged particles.
Also assume a large, but finite magnetization --- that is a large enough magnetic field.

Let observe the motion of the particle between successive collisions.
After a collision all the directions of the velocity are
equiprobable; the mean parallel velocity is zero. Due to action of parallel
Lorentz force there is acceleration
\begin{equation}\label{LorAccel}
a = \langle f \rangle/ m
= \mu \nabla \! B / m=-V^{2}_{\perp} \nabla \! B /2 B.
\end{equation}
Because the electron velocity is larger than
the ions, it is seen from Eq.~(\ref{LorAccel}) that there is the preferred flow
of electrons, that is a current. To evaluate this current,
collisions should be involved. Due to the parallel Lorentz force,
directional velocity arises between collisions. As result
of next collision, this directional velocity again disappears.
To the time $t = 1/\nu$ (mean time between collisions),
the directional velocity attains the value
\begin{equation}
a/ \nu = - \mu \nabla \! B / m \nu .
\end{equation}
The mean (drift) velocity may be written as the half of that:
\begin{equation}
V_{dr} = - \mu \nabla \! B / 2 m \nu .
\end{equation}
For current density the standard formula may be written:
\begin{equation}\label{DriftVel}
{\bf j} = q N {\bf V}_{dr}.
\end{equation}
Here $q$ is the charge, $N$ is the concentration
of charged particles of given sort. Substituting the
need parameters in Eq.~(\ref{DriftVel}), the parallel current density
may be expressed, on module
\begin{equation}\label{ParGrad}
 j_{\parallel} = \frac{q P \nabla \! B}{2 \nu m B}.
\end{equation}
Here $m V^{2}_{\perp}N/2 \equiv P_{\perp} = P$,
$P_{\perp}$ is the perpendicular pressure,
$P$ is the pressure of the charged particles.

Let's compare the parallel currents for electrons and ions
assuming their temperatures being same. Transport cross sections
for the ion-atom collisions exceed them for the electron-atom ones by some
times. Accepting, nevertheless, for simplification, these cross sections
being identical, we receive from Eq.~(\ref{ParGrad}) the approximate dependence of
the parallel current on the mass of the particle
\begin{equation}\label{MassLaw}
j_{\parallel} \, \propto \, 1 / \nu m \, \propto \, m^{-1/2}
\end{equation}
The density of the electron current is very greater than the
ion's. This feature is explicitly shown on Fig.~\ref{FigParal}.
The dependence expressed by Eq.~(\ref{MassLaw}) is same as that for
conductivity current.

\subsection*{Algorithm for calculating the current density}

As mentioned, the density of the Hall gradient current was deduced
for a case of collisionless
plasma~\cite{Alfven:book}. In such the calculations a drift approach
is generally used. This approach has some limitations. Now it is considered
more reliable to use an algorithm suitable for arbitrary magnetization $\beta$.
Equation for the collisionless plasma may be received as the limit
$\beta \rightarrow \infty$~\cite{UFN_Klim}. In the present work we will follow
such a way.

\begin{figure}
\begin{center}
\includegraphics[height=42mm]{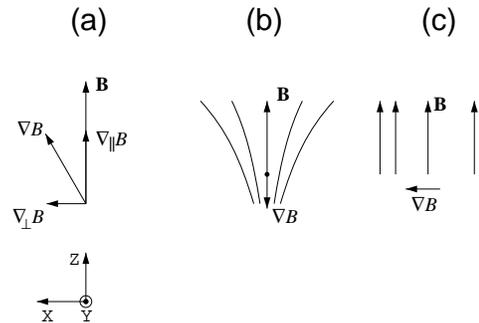}
\end{center}
\caption{Coordinate system and magnetic fields being used
for calculating.
(a) arbitrary disposition of vectors $\nabla B$  and  ${\bf B}$.
Vector $\nabla B$  may be decomposed at parallel and perpendicular
components, which defines, in this work, the axes Z and X respectively.
Vector  ${\bf B}\! \times \! \nabla B$  defines the axis Y. (b) non-uniform magnetic field
with $\nabla B \parallel {\bf B}$. It was used for calculating the parallel component
of gradient current, see also Fig.~\ref{FigParal}. (c)~non-uniform magnetic
field with  $\nabla B \!\perp \!{\bf B}$. It was used for calculating the perpendicular
and Hall components of gradient current, see also Fig.~\ref{FigHall}.
}
\label{FigCoord}
\end{figure}

We took the algorithm what widely exploited for finding the
drift velocity in homogeneous ionized medium with applied
uniform electric and magnetic fields, but we used this algorithm
when non-uniform magnetic field is applied. (In the case of
non-homogeneous medium --- there are gradients of density or
temperature --- another, more complex algorithm should be applied.)

Directly after a collision all the directions of the scattered
particle are equiprobable, --- the collision makes the mean velocity
to equal zero. Then the particle interacts with the field, --- electric,
magnetic or both. This interaction brings asymmetry
in the movement of the particle.
To the time of next (second)
collision the particle has received some average shift in space.
Let ${\bf r}$ be the radius-vector of the particle in the moment of
the next collision relatively the first. Then the average shift
between the subsequent collisions is
\begin{equation}
\label{Rmean1}
\langle{\bf r}\rangle=\sum{{\bf r}_{i}/n_{coll}}.
\end{equation}
Here $n_{coll}$  is the number of the collisions has been included in the calculation.

As a result of the second collision, the directions of the velocity
become again isotropic. Assuming stationarity, the velocity distribution
function does not depend on time. Thus the movement after the second collision
will occur in same a way as after the first. Hence, it is well enough
to take into account the movement of the particle between the first
and second collisions. Inelastic collisions may be omitted in the
calculation because they are already taken into account in the velocity
distribution function. Thus, average (drift) velocity, which the particles
get in the applied field, is
\begin{equation}
{\bf V}_{dr}=\langle{\bf r}\rangle \nu .
\end{equation}
As the drift velocity becomes known, Eq.~(\ref{DriftVel}) gives the current density.

It follows from above, that the calculation of the drift velocity
reduces to calculating the vector $\langle {\bf r} \rangle$.
Several methods --- Monte Carlo, regular, and combined --- were used to yield
an identical result. The regular method as being fastest has been described below.

\begin{figure}[t]
\begin{center}
\includegraphics[height=38mm]{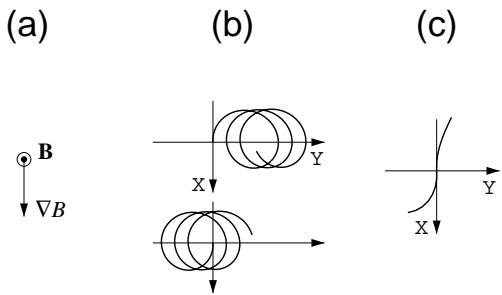}
\end{center}
\caption{
The mechanism of perpendicular and Hall components of gradient current.
(a) the magnetic field being used for calculating. The field growths downwards;
the gyroradius decreases downwards. (b) the origin of drift velocity for
large
magnetization (strong field, sparse plasma). (c) the case of small magnetization.
First collision occurs in the center of coordinate.
The motion of positive particle being scattered up and down is shown; the
trajectories
have the length being equal to free path. Mean shift between collisions is
defined as center of gravity for the locations of next collision.
As obvious from panel (b), for large $\beta$ the mean Y coordinate of second
collision is positive --- this is well-known case of Hall gradient current in collisionless
plasma.
To the contrary,  it is seen from panel (c) that the mean Y coordinate for small $\beta$ is negative.
As consequence,
the Hall component has different signs for small and large
$\beta$, see also Eq.~(\ref{Density16}).
From panel (c) it is seen that a net shift along axis X is present, at its negative
direction. This means that the perpendicular component indeed takes place.
With increasing $\beta$, as seen from panel (b), the mean X coordinate of second collision tends fast
to zero. This well corresponds to behavior of the perpendicular (diagonal) terms in matrix of Eq.~(\ref{Density16}).
}
\label{FigHall}
\end{figure}

The solid angle $4 \pi$,   in which the particle equiprobably scatters,
is divided into small angles $\Omega_{i}$.
The vector of initial velocity is the average on the angle $\Omega_{i}$:
\begin{equation}
{\bf V}_{i}(0) = V(0) \langle {\bf \Omega}_{i} \rangle.
\end{equation}
Here $V(0)$ is the module of initial velocity (for its choice see below),
$\langle {\bf \Omega}_{i} \rangle$ is the mean normal to the area $\Omega_{i}$.
In this method the concept of weight of scattered particle is used.
The weight at the moment immediately after the collision is supposed
$W_{i}(0) = \Omega_{i}/ 4 \pi $, what corresponds to natural normalization
\begin{math}
\sum_{}{}W_{i}(0) = 1
\end{math}.
When the particle is moving on a trajectory
there is reduction of the weight due to collisions. During time $\Delta t$
the weight diminishes by
\begin{displaymath}
\Delta W(t)= W'(t) \Delta t = - W(0) \exp(-t \nu )\Delta t .
\end{displaymath}
The code builds the trajectory of 
the particle moving in given field.
For each the time interval the value $| \Delta W_{i}(t) | {\bf r}_{i}(t) $
is added to $\langle {\bf r} \rangle$.
The calculation for given scattered particle expires when its weight
becomes less than a little value defined beforehand. Finally,
\begin{displaymath}
\langle {\bf r} \rangle = \sum | \Delta W_{i} (t) |
{\bf  r}_{i}(t).
\end{displaymath}
This equation has the same meaning as Eq.\ \ref{Rmean1}.
The sum is taken on the directions of scattered particles,
on intervals of time and, if required, on the module of the initial velocity.

\begin{figure}[t]
\begin{center}
\includegraphics[height=38mm]{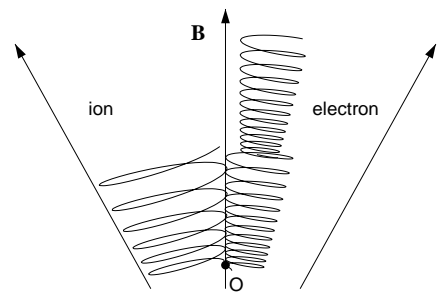}
\end{center}
\caption{The mechanism of parallel gradient current. The motion of electron
and ion in the magnetic field with $\nabla B \parallel {\bf B}$
 is shown between the collisions.
The particles were assumed to have same energies. For clearness, the
ion mass is taken by factor 4 larger than that of electron one. Hence,
the ratio of gyroradiuses is of 2. The particles scattered in point
"O" in the direction perpendicular to the field. First, a trajectory
represents a circle; it turns out into a spiral due to parallel Lorentz
force acting upwards. The trajectories of both particles are depicted for the time
being equal to ion collision time. As for electron, it underwent
two collisions for this time. (Assuming the cross sections are same.)
As seen, the electron has shifted at weaker field direction by two times larger
than the ion. Hence, the drift velocity is inversely proportional to
the root of mass; see Eq.~(\ref{Density16}).
}
\label{FigParal}
\end{figure}

The computation was made for Maxwellian distribution function of charged particles on velocities.
Bulk of the computation has been made for the monoenergetic
particles with the initial velocity equal to the thermal velocity
\begin{displaymath}
V(0)=V_{T} \equiv (8kT/ \pi m)^{1/2}.
\end{displaymath}

Additional computation has been made in which the initial
velocity varied around $V_{T}$;
the initial weight was multiplied by a normalized factor.
It presented the fraction of the particles with the given velocity
module in the Maxwellian distribution. The both computations ---
with fixed and varying velocity module --- gave the same result.

Code execution yields in numerical data of the drift velocity,
being computed for the different magnetization $\beta$.
The code was used for calculation of drift velocity in the fields:

The non-uniform magnetic field for a case  $\nabla \! B \parallel {\bf B}$,
Fig.~\ref{FigCoord}(b) and Fig.~\ref{FigParal}.

The non-uniform magnetic field for a case  $\nabla \! B \perp {\bf B}$, Fig.~\ref{FigCoord}(c)
and Fig.~\ref{FigHall}.

Uniform electric and magnetic fields for a case ${\bf E} \parallel {\bf
B}$.

Uniform electric and magnetic fields for a case ${\bf E} \perp {\bf
B}$.

The last two cases served for the aim of control. They gave the numerical values which corresponds to
a well-known equation of tree-components conductivity current.

\subsection*{Equation for gradient current density}

To fit the numerical data generated by the code,
the empirical equation for gradient current density
was devised. The equation reprodices the numerical data with accuracy of 1\% for a small magnetization
and 5\% for large one:

\begin{equation}\label{Density16}
{\bf j}_{\nabla B}=
-\frac{8qP}{\pi^{2} \nu m L}
\left ( \begin{array}{ccc}
\vspace{1mm}
\frac{2Z\beta^{2}}{(\beta^{2}+1)^{2}}  &  \frac{-\beta(\beta^{2}-1)}{(\beta^{2}+1)^{2}}  & 0 \\
\vspace{1mm}
\frac{\beta(\beta^{2}-1)}{(\beta^{2}+1)^{2}} & \frac{2Z\beta^{2}}{(\beta^{2}+1)^{2}}     &   0 \\
0         & 0                                      &  \frac{Z\beta^{2}}{\beta^{2}+1} \\
\end{array} \right)
{\bf I}.
\end{equation}

Here ${\bf I} $ is the unit vector in direction
of the gradient of magnetic field module,
$L$ is the characteristic length of change of magnetic field
module:
\begin{equation}\label{Lb}
{\bf I} = \nabla \! B/ |\nabla \! B|
, \qquad
L = B / | \nabla \! B|,
\end{equation}

$\nabla \! B \equiv
\left( \begin{array}{c}
\nabla_{X}B    \\
\nabla_{Y}B    \\
\nabla_{||}B    \\
\end{array} \right),$
$Z$ is the sign of charge.

It is seen from Eq.~(\ref{Density16}) that Hall component
of gradient current (being long known for infinite $\beta$)
has the remarkable feature: it changes the direction at  $\beta = 1$.
Fig.~\ref{FigHall} shows why this occurs. This property was not
known formerly.

For comparison, let's copy out from~\cite{N971:Tzen} the equation for the
density of diffusion and thermocurrent for weakly ionized plasma,
for electrons:
\begin{equation}\label{DensityDT}
{\bf j}_{D,T}=
\frac{qP}{\nu m L}
\left ( \begin{array}{ccc}
\vspace{1mm}
\frac{Z}{\beta^{2}+1}  &  \frac{-\beta}{\beta^{2}+1}  &   0 \\
\vspace{1mm}
\frac{\beta}{\beta^{2}+1} & \frac{Z}{\beta^{2}+1}     &   0 \\
0         & 0                                      & -Z \\
\end{array} \right)
{\bf I}.
\end{equation}

If Eq.\ (\ref{DensityDT}) describes the diffusion current,
$$
L \equiv N_{e}/|\nabla \! N_{e} |, \qquad {\bf I} \equiv \nabla N_{e}/|\nabla N_{e}|,
$$
where $L$ is the characteristic length of change of electron density and
${\bf I}$  is the unit vector on concentration gradient.

If Eq.~(\ref{DensityDT}) describes the thermocurrent,
$$
L \equiv T_{e}/|\nabla \! T_{e} |, \qquad {\bf I} \equiv \nabla T_{e}/|\nabla T_{e}|,
$$
where $L$ is the characteristic length of change of electron temperature and
${\bf I}$  is the unit vector on temperature gradient.
For the case of thermocurrent, Eq.~(\ref{DensityDT}) is valid
when collision frequency independent (weakly dependent) of charged particle velocity.

Equation~(\ref{DensityDT}) will also describe the conductivity current if
admit
$L=\epsilon_{T}/qE$ and ${\bf I}={\bf E}/E$, that is L is the length
on which the electron attains the thermal energy
$\epsilon_{T}$ in the given electric field $E$ and
${\bf I}$ is the unit vector in the electric field direction.

It is seen from comparison of equations~(\ref{Density16})
and~(\ref{DensityDT}) that the density
of gradient current at large $\beta$ has the same asymptotic as that for diffusion
and thermocurrent. The parallel components of all the
currents are independent of $\beta$, Hall components behave as $1/ \beta$,
and the perpendicular components behave as $1/ \beta^{2}$.

Equations~(\ref{Density16}) and~(\ref{DensityDT}), being written in the same
form, allow us to evaluate
and to compare the densities of different currents in the plasmas of interest.

\subsection*{Currents of non-uniformities as extraneous currents}

A current may be classified in relation to electric field --- the current
independent on electric field may be extraneous.
For an extraneous current, the factor of dissipation ${\bf j \cdot E}$ is negative.
It is positive for conductivity current and it is zero for Hall current ---
the later is nondissipative current. Due to action of extraneous current a part
of the energy of charged particles turns into the energy of electric and
magnetic field.

Diffusion and thermocurrents, being independent of electric field,
may be extraneous currents --- at proper direction of the electric field.
Generators of electricity have been worked out on base
of diffusion and thermocurrents.

Being also independent of electric
field, gradient and centrifugal currents
may also be  extraneous. A place where they evince as extraneous
is again the magnetosphere. Be the magnetosphere axially symmetric,
the current (sum of gradient and centrifugal ones) were of constant
value along its round. No electric field would arise. But as result
of interaction with solar wind, the magnetosphere is an
asymmetric
structure. Hence the sum of two currents has different value along
equatorial round of the magnetosphere. Is the total (extraneous) current unclosed?
 No, gradient and centrifugal currents create a
charge separation, followed by electric field. Conductivity current
appears. So, total current (including the tree currents) forms
the closed circuit of observed magnetospheric ring
current. (This circuit may be partly branched off in ionosphere.)

\subsection*{Currents of non-uniformities on sun photosphere and on other plasmas}

Let's raise the question, --- currents of what nature participate in
the currents observed in atmosphere of the Sun? For the answer we
will evaluate, with use of equations~(\ref{Density16}) and
(\ref{DensityDT}), the densities of different currents at photosphere
level and compare them with the observable current density.

The electron concentration is taken from~\cite{AA_385:1056}.
From these data, the characteristic length of change
of electron density at vertical has been calculated; it turned out to be
70~km. From the same data, the characteristic length of change
of electron temperature on vertical at photosphere level
was obtained to be about 600~km.
The electron density changes faster than the temperature.
Such the relationship is expected for weakly
(partly) ionized plasma: on Saha equation, a small change in
temperature leads to larger change in degree of ionization,
following in large change in the electron density.

For evaluating the gradient current density with use
Eq.~(\ref{Density16})
we need know the typical length of change of magnetic field
module. Magnetic field in solar atmosphere is known to be
very structural, it rather consists of small-scale magnetic
elements --- magnetic flux tubes~\cite{SP_32:Stenflo}. The cross
size of the magnetic elements  at photospheric level is smaller than the
resolution of magnetometers. Hence it is impossible building
tree-dimensional structure of the magnetic field and defining
the length of change of magnetic field module from it. There are methods
what allow us to learn more about the magnetic field at smaller
scales than the resolution limit.

The procedure of inversion of Stokes profiles is applied to
the polarization of magnetic lines in single resolution
elements~\cite{ApJ_532:1215}. Inversion of Stokes profiles reveals that two
different magnetic components coexist in one resolution element
with the size of 0".3--1".0~=~220--750~km. A large fraction of the
field strengths was measured being in kG regime. A fraction of observed
Stokes profiles requires opposite polarities in the resolution element.
Such the method being especially applied to the different lines or to the
lines in visible and infrared range allows evaluating the mean unsigned
flux density and the longitudinal component of the magnetic field. Then,
filling factor (being one -- several percents) and character size of the
magnetic elements may be obtained~\cite{AA_407:741}.

The next data on characteristic size (diameter) of small-scale magnetic
elements in quiet photosphere may be found in literature:
75~km~\cite{AA_407:741}, 96--118~km~\cite{ApJ_553:449}, 100--200~km~\cite{AA_509:435},
40--220~km~\cite{SP_144:1}, 100~km~\cite{ApJ_463:797}, 50--100~km~\cite{AA_408:1115},
140~km~\cite{Arxiv:org}.

Electron magnetization $\beta$ being calculated for the effective field 1~kG
is given on Tab.~\ref{non-curr}.

Another method for revealing the magnetic structures includes imaging
the Sun in molecular G-band~\cite{ApJ_553:449}, in the wings of strong spectral
lines such as H$\alpha$ and Ca\,{\small II} H and K~\cite{AA_509:435}.
It was stated that the mean
equivalent diameter of the magnetic structures is of 100--300~km, at the limit of
resolution of ground-based solar telescopes.

Summing these data, we may say that the characteristic diameter of small-scale
magnetic elements in quiet photosphere is of 75--300~km. Obviously,
for the characteristic length of change of magnetic field module
the associated radius, 40--150~km, should be taken.

Evidently, there is mechanical balance on horizontal of the magnetic
structures with surrounding plasma~\cite{ApJ_544:1135}. Gas pressure and magnetic
pressure, being defined at the same photospheric level, must anticorrelate
in accordance with equation $P_{gas} + B^{2}/8 \pi = $~const.
In such the case, the typical length of
plasma density on horizontal should be the same as the characteristic
length of change of magnetic field module that is 40--150~km. This well
agrees with the typical length of plasma density on vertical, 70~km.

The given data allows us to evaluate the densities of different currents
on photospheric level. Because the magnetization is close to 1, the
matrix elements  in equations~(\ref{Density16}) and ~(\ref{DensityDT}) is of order 1 also.
Hence, for our aim, either component of some current density
may be presented by reduced equation
\begin{equation}\label{DensityEval}
j \approx q P / \nu m L,
\end{equation}
where the length $L$ accounts for 70, 600, and 40--150~km for diffusion,
thermocurrent, and gradient current, accordingly.

\begin{table}
\caption{\label{non-curr} Electron magnetization $\beta$ and
densities (in mA/m$^{2}$) of currents of non-uniformities at photosphere levels of quiet
sun evaluated on equations (\ref{Density16}) and (\ref{DensityDT}). }
\begin{ruledtabular}
\begin{tabular}{ccddc}
H\enspace(km)  & $\beta$ & j_D & j_T & $j_{\nabla B}$\\
\hline
$-$40  & 0.41  & 6.7 & 0.6 &3.3--13\\
$-$30 & 0.55 & 3.5 & 0.3&1.7--7\\
$-$17  & 0.69 & 1.55 & 0.14 &0.8--3\\
0 & 0.82 & 0.7 & 0.06 & 0.35--1.4\\
22 & 0.97 & 0.4 &0.05 & 0.2--0.8\\
\end{tabular}
\end{ruledtabular}
\end{table}

\begin{table}
\caption{
Densities of vertical current measured at rotor of magnetic field at photosphere
level.}
\begin{ruledtabular}
\begin{tabular}{llc}
j$_z$\enspace(mA/m$^{2}$) & Resolution & Reference      \\ \hline
\multicolumn{3}{c}{ Active regions }\\ \hline
$\pm 2, \ldots \pm 20$  & 2"/pixel  & \cite{AA386:646}  \\
$\pm 7.5,\ldots \pm 12$ & 3" $\times$ 2" & \cite{SP_N1065} \\
$\pm 1.2, \pm 2.4$  & 4" & \cite{SP_N:1081}  \\
$\pm 6, \ldots \pm 24$ & 2" $\times$ 2" & \cite{AA_336:359} \\
$-30, +50$   & 1.1" & \cite{AA392:1119} \\
$\pm 20, \ldots \pm 70$ & & \cite{SP_N1071}  \\
$\pm 7 $ & & \cite{SP174:175} \\
$\pm 2.4, \ldots \pm 20$ &   & \cite{AA_N:1057} \\ \hline
\multicolumn{3}{c}{ Quiet photosphera } \\ \hline
$\pm 0.1, \ldots \pm 0.2$ & 4" &  \cite{419b} \\ 
\end{tabular}
\end{ruledtabular}
\label{vert-curr}
\end{table}

The densities of currents of non-uniformities, being calculated at Eq.~(\ref{DensityEval}) for
these lengths are given on Tab.~\ref{non-curr}.

Now, of concern to us is observable currents in sun photosphere. The only current what
can be measured in there is magnetic field-aligned current in areas
of strong vertical field. Here $\nabla \! B \! \parallel \! {\bf B}$.
In active regions including sunspots this current may be measured with enough accuracy
on rotor of cross-to-sight magnetic field. See Tab.~\ref{vert-curr} for the data.
It is seen that the better the resolution, the larger is the
measured value of current density.
This correlation was many discussed in literature. (This particularity in
measurements appears because the current/magnetic field structures have
sizes smaller than the resolution; and the currents in close structures can
have opposite direction.)

It is difficult to measure transverse magnetic field in quiet sun photosphere and
to calculate vertical current there. Nevertheless, the estimation of the
quiet sun currents was made with use of chromospheric fine structures seen in
H$\alpha$ images with comparison of theoretical chromospheric field obtained by
potential extrapolation of the observed, line-of-sight photospheric
field~\cite{SP_N1072,419b}.
In area of unipolar vertical magnetic field, there
were found the currents (1--8)$\times$10$^9$A being directed upwards or
downwards, with the density of 0.1--0.2 mA/m$^{2}$~\cite{419b}.
This density seems to be underestimated when the low
resolution in this work is considered.
The structure of currents in unipolar hill of vertical field in quiet regions
resembles the current structure in a sunspot.

We want to compare the measured currents with the currents of
non-uniformities; in this comparison the currents must fall
to region with same conditions --- quiet photosphere and,
separately, active regions.

{\it Quiet photosphere.} Let us compare the density of currents of
non-uniformities presented in Tab.~\ref{non-curr} with only measured
current presented in Tab.~\ref{vert-curr}.
We may see that the measured
current is smaller (even with regard to its underestimation)
than the diffusion and gradient ones.
From this it follows that the currents of non-uniformities are present in
quiet photosphere and play important role in there.

{\it Active regions.} It is known that both magnetic fields and
gradients (of concentration, temperature, and magnetic field module)
are larger in active photosphere than in quiet one. Hence, we may
anticipate, that the currents of non-uniformities in active regions
are larger than in quiet photosphere. Again, we see that the currents
of non-uniformities participate in the photosphere currents and play
essential role.

To discuss the problem further, let us take in mind the directions of the currents.
Because $\nabla \! B \! \parallel \! {\bf B}$, as the result of given work there
is the parallel gradient current there; it consists of electrons
moving upwards, to weaker magnetic field region. There is the parallel diffusion current
there, it also consists of electrons moving upwards, to smaller electron
concentration. Thus, both currents act together. They act as the extraneous
current and give a charge separation following with macroscopic electric
field and conductivity current. The later is of opposite direction to
the extraneous current. It may be proposed that there is difference of
extraneous and conductivity
currents; this difference is directed upwards or downwards.

In Earth polar ionosphere, there are seen phenomena being considered
as of similar nature to paired vertical currents on solar atmosphere.
These are U-shape structures consisting
of two magnetic field-aligned currents of opposite directions on distance
of some hundred kilometers from each other~\cite{U_shape}.
It is appropriate mention here that both solar and ionospheric
vertical currents are closed in area where electron magnetization is near~1.

Phenomena accompanying the magnetic field-aligned gradient current are
observed in laboratory plasma. The plasma facility described
in~\cite{Ambal} has a magnetic field with axial symmetry.
The plasma source  is located where the field is maximal. In the center
of the plasma volume the magnetic field is weakest; and the negative
potential about $- 300\,$V has been measured there. On frame of the given work
it is possible to assume that parallel gradient plus diffusion currents
cause the charge separation in the plasma; the negative potential evidences it.
Also, a through current $\approx 1\,$kA streaming along magnetic
lines is observed in the plasma. The direction of this current corresponds to
movement of electrons towards weaker magnetic field and, at the same
time, towards weaker electron density. The through current is presumably
the gradient plus diffusion current, being partly canceled by the conductivity
current of opposite direction.

In the plasmas of interest the cross-magnetic field currents are observed:
1) the current in photosphere which closes the vertical currents inside
a sunspot and inside the hill of vertical field in quiet region,
2) the current in lower ionosphere what closes magnetic
field-aligned currents in U-shape aurora structure, 3) equatorial
electrojet in lower ionosphere at the height 90--110~km. Presumably,
cross components (these are perpendicular and Hall ones) of gradient,
centrifugal, diffusion, and also of conductivity current include to these
observable currents.
Notice again that magnetization of electrons is near~1 in the areas of these
transverse-magnetic field currents.

Extraneous currents (gradient, centrifugal, diffusion, and thermocurrent)
when being coupled with
conductivity current may be considered of to generate the magnetic
fields in objects without convective motions --- in stars of classes Am, Ap.

\subsection*{Summary}
Results of this paper may be summarized as the follows.
\begin{enumerate}
\item

We gathered the observations of electrical currents in cosmic
plasmas paying attention to the cases when nature of the current
is known. By this is meant that the value of measured current
agrees sufficiently with the current calculated on a theoretical equation
(for a current of some type).
The nature of the ring current
in magnetosphere is well known --- this is gradient plus centrifugal current.
The current in comet coma and in artificial plasma cloud consists
of diffusion and thermocurrent. These four currents are the current of
non-uniformities.
\item
In the paper the gradient current was investigated more fully than
earlier. Two unknown components of the current were found. An
equation for gradient current density was obtained. The equation
embraces tree possible components of the gradient current.
\item
It was proposed that the currents of non-uniformities present on
solar atmosphere. It was checked for photosphere. The measured
vertical currents was compared with the currents of
non-uniformities; the later were calculated at theoretical
equations. Theoretical values turned out to be the same or larger
than the measured ones.
\item
Hence it follows that the currents of non-uniformities indeed present
in atmosphere of the star; the currents should be included into
theoretical models.

\end{enumerate}


\clearpage

\begin{thebibliography}{1}

\bibitem{PR_61:N836}
S.~I.~Vainshtein, S.~M.~Chitre, and A.~V.~Olinto,
\pre {\bf 61}, 4422 (2000).

\bibitem{PRL_93:N1062} J.~D.~Huba and L.~I.~Rudakov,
\prl {\bf 93}, 175003 (2004).

\bibitem{PhScr_57:N892} Chang-Mo Ryu and M.~Y.~Yu,
Phys. Scr. {\bf 57}, 601 (1998).

\bibitem{PRL_N:726}
A.~B.~Hassam, R.~M.~Kulsrud, R.~J.~Goldston {\it et al.},
\prl {\bf 83}, 2969 (1999).

\bibitem{ApJ_1034:509}
S.~J.~Desch,
\apj {\bf 608},  509 (2004).

\bibitem{SP_N:1084}
I.~J.~D.~Craig and P.~G.~Watson,
Sol. Phys. {\bf 214},  131 (2003).

\bibitem{Severny1965}
A.~B.~Severny, Space Sci. Rev. {\bf 3}, 451 (1965).

\bibitem{SP_N:900Abram} V.~A.~Abramenko,
S.~I.~Gopasyuk, M.~B.~Ogir,
Sol. Phys. {\bf 134}, 287 (1991).

\bibitem{AA_N:1057} Hongqi Zhang,
Astron. Astrophys. {\bf 324}, 713 (1997).

\bibitem{SP_N:1081} V.~M.~Grigoryev and L.~V.~Ermakova,
Sol. Phys. {\bf 207}, 309 (2002).

\bibitem{SP_N:1069}Yang Liu, Maki Akioka, Yihuayan and Guoxiang Ai,
Sol. Phys. {\bf 177}, 395 (1998).

\bibitem{AA386:646} Y.~Liu and H.~Q.~Zhang,
Astron. Astrophys. {\bf 386}, 646 (2002).

\bibitem{SP174:175}
Jeongwoo Lee, Stephen M.~White, N.~Gopalswamy, and M.~R.~Kundu,
Sol. Phys. {\bf 174}, 175 (1997).

\bibitem{AA392:1119}
S.~R\'egnier, T.~Amari, and E.~Kersal\'e,
Astron. Astrophys. {\bf 392}, 1119 (2002).

\bibitem{AA_337} V.~V.~Zaitsev, A.~V.~Stepanov, S.~Urpo, and S.~Pohjolainen,
Astron. Astrophys. {\bf 337},  887 (1998).

\bibitem{AA_351:Titov} V.~S.~Titov and P.~Demoulin,
Astron. Astrophys. {\bf 351}, 707 (1999).

\bibitem{AA_336:359}
Wang Tongjiang, Qiu Jiong, and Zhang Hongqi,
Astron. Astrophys. {\bf 336}, 359 (1998).

\bibitem{ApJ_557:880} M.~R.~Kundu, A.~Nindos, S.~M.~White, and V.~V.~Grechnev,
\apj {\bf 557}, 880 (2001).

\bibitem{SP_N1071} H.~ S.~Ji, M.~T.~Song, X.~Q.~Li and F.~M.~Hu.
Sol. Phys. {\bf 182}, 365 (1998).

\bibitem{SP_N1077}V.~G.~Lozitsky, E.~A.~Baranovsky, N.~I.~Lozitska, and
U.~M.~Leiko,
Sol. Phys. {\bf 191}, 171 (2000).

\bibitem{Jacobsen} G.~Jacobsen , R.~Carlqvist,
Icarus {\bf 3}, 270 (1964).

\bibitem{SP_1:Alfven}H.~Alfven and R.~Carlqvist,
Sol. Phys. {\bf 1}, 220 (1967).

\bibitem{SP_N1065} Unwei Zhao, Cheng Fang, and Ming de~Ding,
Sol. Phys. {\bf 173}, 121 (1997).

\bibitem{SP_N1080} I.~A.~Bilenko, A.~I.~Podgorny and
I.~M.~Podgorny,
Sol. Phys. {\bf 207}, 323 (2002).

\bibitem{SP_N1066} Yoichiro Hanaoka,
Sol. Phys. {\bf 173}, 319 (1997).

\bibitem{AA_344:981}R.~Falewicz and P.~Rudawy,
Astron. Astrophys. {\bf 344}, 981 (1999)

\bibitem{AA_334:L57}
V.~I.~Abramenko, V.~B.~Yurchishin, and V.~Carbone,
Astron. Astrophys. {\bf 334}, L57 (1998).

\bibitem{AA_348:405} J.~L.~Han, R.~Beck, M.~Ehle, R.~F.~Haynes,
and R.~Wielebinski {\it et al.},
Astron. Astrophys. {\bf 348},  405 (1999).

\bibitem{Parker:Book} E.~N.~Parker, {\it Cosmical Magnetic Fields},
(Clarendon Press, Oxford, 1979).

\bibitem{AA_344:923}C.~Thum and D.~Morris,
Astron. Astrophys. {\bf 344}, 923 (1999).

\bibitem{DAN_N464}S.~S.~Dolginov, N.~V.~Pushkov,
Dokl. Akad. Nauk USSR {\bf 129}, 77 (1959).

\bibitem{UFN_N220} V.~I.~Belinicher and B.~I.~Sturman.
Usp. Phiz. Nauk {\bf  130}, 415 (1980).

\bibitem{N971:Tzen} V.~A.~Rozhansky, L.~D.~Tsendin,
{\it Transport Phenomena in Partially-Ionized Plasmas},
(Bristol, Gordone \& Breach, 2000).

\bibitem{AA_N1052}P.~L.~Israelevich, A.~I.~Ershkovich, and F.~M.~Neubauer,
Astron. Astrophys. {\bf 329}, 765 (1998).

\bibitem{GRL_N786} B.~G.~Gavrilov,
A.~I.~Podgorny, I.~M.~Podgorny {\it et al.},
Geophys. Res. Lett. {\bf 26}, 1549 (1999).

\bibitem{PRL_93:N1047} D.~M.~Thomas, A.~W.~Leonard, L.~L.~Lao, T.~H.~Osborne,
\prl {\bf 93}, 065003 (2004);
M.~R.~Wade, M.~Murakami, and P.~A.~Politzer,
\prl {\bf 92}, 235005 (2004).

\bibitem{Raizer} Yu.~P.~ Raizer,
{\it Gas Discharge Physics}, (Springer, Berlin, 1991).

\bibitem{Alfven:book} H.~Alfven, C.-G.~Falthammar, {\it Cosmic Electrodynamic},
(Oxford, Clarendon Press, 1963, Second Edition).

\bibitem{UFN_Klim} Yu.~L.~Klimontovich,
Physics-Uspekhi {\bf 40}, 21 (1997).

\bibitem{AA_385:1056} J.~M.~Borrero1 and L.~R.~Bellot Rubio,
Astron. Astrophys. {\bf 385}, 1056 (2002).

\bibitem{SP_32:Stenflo}
J.~O.~Stenflo, Sol. Phys. {\bf 32}, 41 (1973).

\bibitem{ApJ_532:1215} J.~S\'anchez Almeida and B.~W.~Lites,
\apj {\bf 532}, 1215 (2000).

\bibitem{AA_407:741} I.~Dom\'\i nguez Cerde\~na, J.~S\'anchez Almeida, F.~Kneer,
Astron. Astrophys. {\bf 407}, 741 (2003).

\bibitem{ApJ_553:449}T.~E.~Berger and A.~M.~Title,
\apj {\bf 553}, 449 (2001).

\bibitem{AA_509:435}
A.~A.~van~Ballegooijen, P.~Nisenson, R.~W.~Noyes {\it et al.},
Astron. Astrophys. {\bf 509}, 435 (1998).

\bibitem{SP_144:1} Z.~Yi and O.~Engvold,
Sol. Phys. {\bf 144}, 1 (1993).

\bibitem{ApJ_463:797} A.~M.~Title and  T.~E.~Berger,
\apj {\bf 463}, 797 (1996).

\bibitem{AA_408:1115}
E.~V.~Khomenko, M.~Collados, S.~K.~Solanki, A.~Lagg and J.~Trujillo Bueno,
Astron. Astrophys. {\bf 408}, 1115 (2003).

\bibitem{Arxiv:org}Itahiza Dominguez Cerdena, Jorge S\'anchez Almeida, Franz Kneer,
astro-ph/0604381 (2006).

\bibitem{ApJ_544:1135} S.~Almeida,
\apj  {\bf 544}, 1135 (2000).

\bibitem{SP_N1072}M.~F.~Woodard, Jongchul Chae,
Sol. Phys. {\bf 184}, 239 (1999).

\bibitem{419b} V.~I.~Abramenko, S.~I.~Gopasyuk, and M.~B.~Ogir.
in {\it Cosmic Researchs}, edited by G.~E.~Kocharov, (Ioffe FTI, S.~Petersburg,
1991), p. 192.

\bibitem{U_shape}
F.~S.~Mozer, C.~W.~Carlson, M.~K.~Hudson {\it et al.},
\prl {\bf 38}, 292 ( 1977);
F.~S.~Mozer, R.~Ergun, M.~Temerin {\it et al.},
\prl {\bf 79}, 1281 (1997).
R.~E.~Ergum, C.~W.~Carlson, J.~P.~McFadden {\bf et al.},
\prl {\bf 81}, 826 ( 1998);
G.~T.~Marklund, N.~Ivchenko, T.~Karlsson {\it et al.},
\nat {\bf 414}, 724 (2001).

\bibitem{Ambal}
T.~D.~Akhmetov, V.~S.~Belkin, E.~D.~Bender {\it et al.},
Plasma Phys. Rep. {\bf 23},  911 ( 1997);
T.~D.~Akhmetov, V.~I.~Davydenko, A.~A.~Kabantsev {\it et al.},
Plasma Phys. Rep. {\bf 24}, 1000 (1998).

\end{thebibliography}
\end{document}